\documentclass[10pt,conference]{IEEEtran}
\IEEEoverridecommandlockouts

\usepackage{booktabs} 
\usepackage{colortbl}
\usepackage{adjustbox}
\usepackage{algorithm}
\usepackage{algorithmicx}
\usepackage{algpseudocode}
\usepackage{amsmath}
\usepackage{amssymb}
\usepackage{bm}
\usepackage{enumitem}
\usepackage{soul}
\usepackage{threeparttable}  
\usepackage{multirow}
\usepackage{makecell}
\usepackage{tabularx}
\usepackage[T1]{fontenc}
\usepackage{array}
\usepackage{listings}
\usepackage{xspace} 
\usepackage{subfigure}
\usepackage[hidelinks]{hyperref}
\usepackage{textcomp}
\usepackage{microtype} 
\usepackage{calc}
\usepackage{pifont}
\usepackage{tcolorbox}
\usepackage{cite}
\usepackage{graphicx}
\usepackage{balance}

\newlength{\ColorBoxDepthReference}
\newlength{\ColorBoxHeightReference}
\newlength{\Width}%
\newcommand{\MyColorBox}[2][red]%
{%
	\settowidth{\Width}{#2}%
	\colorbox{#1}%
	{%
		\raisebox{-\ColorBoxDepthReference}%
		{%
			\parbox[b][\ColorBoxHeightReference+\ColorBoxDepthReference][c]{\Width}{\centering#2}%
		}%
	}%
}
\newcommand\orangebox[1]{\MyColorBox[orange!30]{#1}}

\definecolor{codegreen}{rgb}{0,0.6,0}
\definecolor{codegray}{rgb}{0.5,0.5,0.5}
\definecolor{codepurple}{rgb}{0.58,0,0.82}

\newcommand{\inlinecode}[1]{``\texttt{\small #1}''}

\newcommand{\first}[1]{\fboxsep1.5pt\colorbox{orange!40}{#1}}
\newcommand{\second}[1]{\fboxsep1.5pt\colorbox{yellow!40}{#1}}
\newcommand{\other}[1]{\fboxsep1.5pt\colorbox{white!0}{#1}}

\newcommand{\summary}[1]{
	\begin{center}
		\begin{tcolorbox}[colback=gray!10,colframe=black!25,width=1\columnwidth,arc=1mm, auto outer arc,boxrule=0.5pt,boxsep=5pt,left=3pt,right=3pt,top=0pt,bottom=0pt]
		\textbf{SUMMARY:} #1
		\end{tcolorbox}
	\end{center}
}

\lstdefinestyle{python}{
    language=Python,
    backgroundcolor=\color{white},
    basicstyle=\ttfamily\footnotesize,
    breaklines=true,
    numbers=left,
    numberstyle=\tiny\color{gray},
    keywordstyle=\color{blue},
    commentstyle=\color{green},
    stringstyle=\color{red},
}

\newcommand{\eg}{\textit{e.g.,}\xspace}
\newcommand{\ie}{\textit{i.e.,}\xspace}
\newcommand{\etal}{\textit{et al.}\xspace}
\newcommand{\cf}{\textit{cf.}\xspace}

\newcommand{\gtr}{\emph{GTR}\xspace}
\newcommand{\app}{\textsc{TIGER}\xspace}
\newcommand{\typegen}{\text{TypeGen}\xspace}
\newcommand{\typewriter}{\text{TypeWriter}\xspace}
\newcommand{\typeforpy}{\text{Type4Py}\xspace}

\newcommand{\mask}{\texttt{\small <TYPE>}}
\newcommand{\bos}{\texttt{\small <BOS>}}
\newcommand{\eos}{\texttt{\small <EOS>}}

\begin{document}
\title{\app: A Generating-Then-Ranking Framework for Practical Python Type Inference}

\author{
    \IEEEauthorblockN{Chong Wang\IEEEauthorrefmark{1}, Jian Zhang\IEEEauthorrefmark{1}\IEEEauthorrefmark{2}, Yiling Lou\IEEEauthorrefmark{3}, 
    Mingwei Liu\IEEEauthorrefmark{4}, Weisong Sun\IEEEauthorrefmark{1}, Yang Liu\IEEEauthorrefmark{1}, and Xin Peng\IEEEauthorrefmark{3}}
    \IEEEauthorblockA{
        \IEEEauthorrefmark{1}\textit{College of Computing and Data Science, Nanyang Technological University, Singapore}\\
        \{chong.wang, jian\_zhang, weisong.sun, yangliu\}@ntu.edu.sg\\
        \IEEEauthorrefmark{3}\textit{School of Computer Science and Shanghai Key Laboratory of Data Science, Fudan University, China}\\ 
        \{yilinglou, pengxin\}@fudan.edu.cn\\
        \IEEEauthorrefmark{4}\textit{School of Software Engineering, Sun Yat-sen University, China}\\
        liumw26@mail.sysu.edu.cn
    }
    \thanks{\IEEEauthorrefmark{2} Corresponding author: Jian Zhang (jian\_zhang@ntu.edu.sg).}
}

\maketitle

\begin{abstract}
    Python's dynamic typing system offers flexibility and expressiveness but can lead to type-related errors, prompting the need for automated type inference to enhance type hinting.
    While existing learning-based approaches show promising inference accuracy, they struggle with practical challenges in comprehensively handling various types, including complex parameterized types and (unseen) user-defined types.

    In this paper, we introduce \app, a two-stage \emph{generating-then-ranking} (\gtr) framework, designed to effectively handle Python's diverse type categories. \app leverages fine-tuned pre-trained code models to train a generative model with a span masking objective and a similarity model with a contrastive training objective. This approach allows \app to generate a wide range of type candidates, including complex parameterized types in the generating stage, and accurately rank them with user-defined types in the ranking stage. Our evaluation on the ManyTypes4Py dataset shows \app's advantage over existing methods in various type categories, notably improving accuracy in inferring user-defined and unseen types by 11.2\% and 20.1\% respectively in Top-5 Exact Match. Moreover, the experimental results not only demonstrate \app's superior performance and efficiency, but also underscore the significance of its generating and ranking stages in enhancing automated type inference.
\end{abstract}

\begin{IEEEkeywords}
type inference, pre-trained code models, generating-then-ranking, contrastive learning
\end{IEEEkeywords}

\maketitle

\section{Introduction}\label{sec:introduction}
Python's dynamic typing system, while fostering adaptable and expressive code, can pose challenges like code comprehension, maintainability, and type-related errors~\cite{pldi/AllamanisBDG20,icse/MirLPG22,kbse/PengWWGL23}.
Python Enhancement Proposals (PEPs)~\cite{PEP484,PEP526} were introduced to enhance Python's type hinting system, advocating for type hints to improve code quality and support tools like static type checking~\cite{Mypy,Pyre,Pyright,pytype}. Despite this, manual type annotation can be labor-intensive and error-prone~\cite{kbse/OreEDK18}.

To this end, there is a growing trend towards adopting automated approaches for inferring variable types through contextual analysis, encompassing both rule-based and learning-based methodologies. Rule-based approaches~\cite{Mypy,Pyre,Pyright,pytype} rely on predefined patterns and rules for accurate type hinting but may lack comprehensive coverage for diverse scenarios~\cite{sigsoft/JesseDA21,kbse/PengWWGL23}.
Learning-based approaches, gaining prominence, transform type inference into different tasks. They can be classified into three types: 1) \textit{Classification-based}. Utilize classification models trained on contextual features to predict variable types~\cite{sigsoft/PradelGL020,sigsoft/JesseDA21,icse/YanFFX23}. 2) \textit{Similarity-based}: Utilize deep learning similarity calculation models to assess candidate types based on their similarity scores, ranking them accordingly. 3) \textit{Generation-based}: Utilize generative models to produce types based on given inputs~\cite{emnlp/0034WJH21,acl/GuoLDW0022,iclr/FriedAL0WSZYZL23,kbse/PengWWGL23}.

However, these learning-based approaches \emph{face limitations in considering various visible types, notably parameterized types and (unseen) user-defined types.} Note that user-defined types in this paper also include third-party types.
Classification-based approaches~\cite{sigsoft/PradelGL020,sigsoft/JesseDA21,icse/YanFFX23} rely on fixed-sized type vocabularies (\eg 1,000 types for TypeWriter~\cite{sigsoft/PradelGL020}), leading to numerous out-of-vocabulary (OOV) types, such as complex parameterized types (\eg \inlinecode{Dict[str,List[int]]})  or unseen user-defined types. Similarity-based approaches~\cite{pldi/AllamanisBDG20,icse/MirLPG22,icse/PengGLG0ZL22} use import information to gather user-defined types and rank candidate types based on contextual similarities. However, these approaches struggle to cover the myriad parameterized types (\eg \inlinecode{Union[str,int]}) formed by elementary types (\eg \inlinecode{Union}, \inlinecode{str}, \inlinecode{int}). Generation-based approaches~\cite{emnlp/0034WJH21,acl/GuoLDW0022,iclr/FriedAL0WSZYZL23,kbse/PengWWGL23} can generate creative types, including complex parameterized types, but most struggle with recognizing and predicting unseen user-defined types not in the training set.

Large language models (LLMs) such as ChatGPT are increasingly utilized for type inference tasks. For instance, Peng et al.~\cite{kbse/PengWWGL23} introduced \typegen, a generation-based approach that utilizes parameter-frozen LLMs and incorporates visible user-defined types in prompt construction. However, it faces accuracy challenges for function arguments as discussed in their paper~\cite{kbse/PengWWGL23}, due to intricacies and robustness issues  with prompt engineering on non-tunable, parameter-frozen LLMs~\cite{chi/Zamfirescu-Pereira23}. \emph{Furthermore, ChatGPT-based approaches, including \typegen, face practical limitations regarding inference cost and scalability.} \typegen relies on external APIs~\cite{OpenAIAPI} for interactions with ChatGPT and employs voting mechanisms~\cite{iclr/0002WSLCNCZ23} for accurate inference, resulting in significant time and financial overhead.

\begin{figure}
    \centering
    \includegraphics[width=1.0\columnwidth]{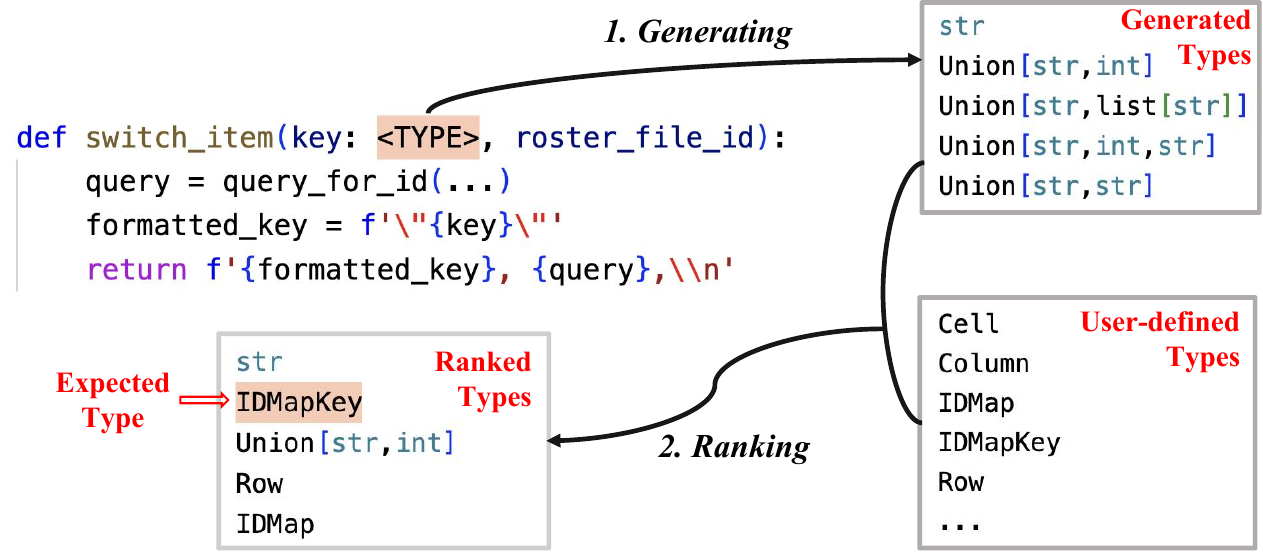}
    \caption{The \emph{Generating-Then-Ranking} Framework}
    \label{fig:framework}
\end{figure}

To overcome these limitations, we introduce a two-stage framework named \emph{generating-then-ranking} (\gtr framework), as depicted in Figure~\ref{fig:framework}. \emph{This framework combines the strengths of both generation-based and similarity-based approaches, allowing for the consideration of comprehensive visible types}. In the initial \emph{generating} stage, our objective is to produce diverse candidate types, including commonly used elementary types like \inlinecode{UUID} and parameterized types like \inlinecode{Union[str,int]}, to fill the type placeholders (\ie \inlinecode{\orangebox{\mask}} in Figure~\ref{fig:framework}). Subsequently, in the \emph{ranking} stage, visible user-defined types, potentially unseen during training, such as \inlinecode{IDMap} and \inlinecode{IDMapKey}, are effectively ranked alongside the generated candidates based on similarities. 
Furthermore, for efficiency and cost considerations, we opt for relatively lightweight pre-trained code models like the 220M-parameter CodeT5~\cite{emnlp/0034WJH21}, specifically fine-tuned for type inference, diverging from the use of LLMs like ChatGPT.

We present \app, a novel \underline{T}ype \underline{I}nference approach that implements the \underline{GE}nerating-Then-\underline{R}anking framework. 
To facilitate the online \gtr inference depicted in Figure~\ref{fig:framework}, \app initiates with offline training of a generation model and a similarity model with collected annotated Python functions. During the generation model training, \app masks annotated types with type placeholders to create type-missed functions. These functions are employed with the span masking objective~\cite{emnlp/FengGTDFGS0LJZ20} to train the model in generating types to fill the placeholders. For training the similarity model, \app analyzes import statements to identify visible user-defined types, which are combined with the types generated by the generation model to form a set of candidate types. Subsequently, a contrastive training objective~\cite{icml/ChenK0H20,corr/abs-1807-03748,nips/OordVK17,cvpr/He0WXG20} is utilized to train the model to learn robust representations for both type-missed functions and candidate types~\cite{nips/AnFLKQ022}, facilitating effective measurement of contextual similarities between them. During online inference, armed with the trained models, \app performs the two-stage generating-then-ranking process. Given a type-missed function containing a type placeholder, this process generates type candidates using the generation model. These are then ranked based on their similarity to user-defined types visible in the scope, using the similarity model.


We extensively evaluate \app's inference effectiveness and efficiency on the ManyTypes4Py dataset~\cite{msr/MirLG21}, a widely adopted benchmark in previous studies. Our results demonstrate \app's superiority in inference accuracy across all type categories compared to six baseline models. Specifically, \app achieves an impressive accuracy rate of 94.1\% and 85.0\% in Top-5 Exact Match for predicting user-defined and unseen types, respectively, marking significant improvements of 11.2\% and 20.1\% over the best baseline approach (RQ1). Furthermore, \app exhibits enhanced robustness across various variable categories, encompassing local variables, function arguments, and return values (RQ2). We also assess the efficiency of \app, highlighting its potential for optimization in large-scale inference scenarios (RQ3). Finally, our ablation study validates the significant contributions of both the generating and ranking stages to \app's overall 
 performance (RQ4).

To summarize, this paper makes the following contributions:
\begin{itemize}[leftmargin=15pt]
    \item A novel two-stage framework, ``Generating-Then-Ranking'', which systematically generates and ranks candidate types alongside user-defined types, facilitating a comprehensive consideration of visible candidate types.
    \item \app, a practical Python type inference approach that implements the two-stage framework through fine-tuning pre-trained code models with specific training objectives.
    \item Extensive experiments on the  ManyTypes4Py benchmark demonstrate the effectiveness and robustness of \app across diverse type and variable categories, particularly for (unseen) user-defined types prevalent in real-world scenarios. Additionally, efficiency evaluations of \app showcase its potential for optimization in large-scale inference scenarios.
\end{itemize}

\section{Related Work}\label{sec:related}

\subsection{Rule-based Type Inference}
Rule-based methods in type inference rely on predefined rules for inferring variable types. A rule can only be activated when all the underlying premises are known, after which it determines the type of the variable based on a set conclusion~\cite{kbse/PengWWGL23}. To address the imperative for static type hints within dynamically typed programming languages, numerous techniques have emerged for the purpose of type inference and verification. Notable examples include Pyright and Pylance (Microsoft)~\cite{Pyright}, Pyre (Meta)~\cite{Pyre}, pytype (Google)~\cite{pytype}, and Python's official type checker, Mypy~\cite{Mypy}.
Apart from industry tools, several academic methodologies have emerged for type inference across various programming languages like Python and JavaScript~\cite{DBLP:conf/ecoop/AndersonGD05,DBLP:conf/popl/0008E16,DBLP:conf/popl/EmmiE16,DBLP:conf/sas/JensenMT09,DBLP:journals/pacmpl/PavlinovicSW21,sigsoft/XuZCPX16}. 
While these approaches excel in terms of accuracy, they confront a significant challenge stemming from dynamic language features and external function calls~\cite{icse/PengGLG0ZL22}. This challenge manifests as a limitation in their coverage, hindering their ability to provide comprehensive type inference scenarios.

\subsection{Learning-based Type Inference}
\emph{\textbf{Classification-based Approaches.}}
In addressing the type inference problem, supervised learning has been employed by researchers with large-scale training data to train classification-based models. Pradel \etal~\cite{sigsoft/PradelGL020} introduced \typewriter, which utilizes probabilistic type prediction and search-based refinement to infer function types from partially annotated code and validate them using a gradual type checker.  Wei \etal~\cite{iclr/WeiGDD20} introduced a probabilistic type inference approach for TypeScript that employs a graph neural network to analyze a program's type dependency graph, facilitating predictions of both standard and user-defined types. Yan \etal~\cite{icse/YanFFX23} proposed DLInfer, an approach that collects slice statements for variables through static analysis and employs a bi-directional gated recurrent unit (GRU) model to learn type propagation information for inference. Additionally, TypeBert~\cite{sigsoft/JesseDA21} demonstrated that by harnessing the token-sequence inductive bias found in BERT-style models and having access to ample data, it is feasible to surpass the type-annotation performance of even the most advanced models.
These approaches encounter practical challenges when dealing with OOV types, including unseen user or library-defined types. As a result, they struggle to accurately predict types for a significant portion of variables in real-world projects.

\emph{\textbf{Similarity-based Approaches.}}
Allamanis \etal~\cite{pldi/AllamanisBDG20} presented a graph neural network model that leverages probabilistic reasoning and deep similarity learning to predict types, including rare and user-defined types, based on a program's structure, names, and patterns. Mir \etal~\cite{icse/MirLPG22} proposed \typeforpy, a hierarchical neural network model employing deep similarity learning to infer likely types for program elements by distinguishing between similar and dissimilar types in a high-dimensional space. Peng \etal introduced HiTyper, a hybrid type inference approach combining static inference and similarity-based models by leveraging type dependency graphs (TDGs) to record and integrate type dependencies among variables, facilitating iterative static inference and neural predictions until the complete inference of TDGs.
While these approaches demonstrate the capability to handle arbitrary types, they require pre-determined candidates, limiting their effectiveness in adequately considering numerous parameterized types.

\emph{\textbf{Generation-based Approaches.}}
Following the type annotation conventions recommended in PEPs, the Python type inference problem can be transformed into a conditional generation problem. For instance, it can be treated as a cloze-style fill-in-blank task where a type placeholder is inserted after a variable that requires annotation. 
Pre-trained code models~\cite{emnlp/FengGTDFGS0LJZ20,emnlp/0034WJH21,acl/GuoLDW0022,iclr/FriedAL0WSZYZL23} are then employed to generate types fill in the placeholder with types based on their learned code naturalness. To fully leverage their generation capability, some studies~\cite{iclr/WeiDD23} have fine-tuned these pre-trained models to improve inference accuracy. 
These generation-based approaches, with fill-in-blank task formulation, lack awareness of unseen user or library-defined types.
In a recent study, Peng \etal~\cite{kbse/PengWWGL23} introduced a generation-based approach known as \typegen for Python type inference, which relies on parameter-forzen LLMs like \texttt{gpt-3.5} and \texttt{gpt-4}. \typegen integrates type dependencies derived from lightweight static analysis with in-context learning~\cite{corr/abs-2301-00234} to construct few-shot Chain-of-Thought (CoT) prompts. These prompts are employed to generate both type predictions and accompanying explanations. 
While this LLM-based generative approach demonstrates notable effectiveness in type inference, it faces practical challenges, particularly with regard to the scalability concerns arising from the resource-intensive interactions with expensive LLMs.

\subsection{Pre-trained Code Models in SE}
Recent research~\cite{icse/WanZZSXJ22,emnlp/0034WJH21,kbse/KarmakarR21} indicates that pre-trained code models encapsulate valuable information about code syntax and semantics, identifier and namespace concepts, and natural language naming. To harness this rich code information for downstream software engineering tasks, researchers commonly fine-tune these models on task-specific data. Examples of such tasks include code generation/completion~\cite{corr/abs-2401-06391,wang2024rlcoder,liu2024stall+,du2024evaluating}, code search~\cite{shi2023cocosoda,corr/abs-1909-09436}, defect detection~\cite{corr/abs-2311-04448}, vulnerability detection/location~\cite{zhang2024empirical}, and program repair~\cite{sp/PearceTAKD23,huang2023empirical}.
Researchers have explored the application of prompt-based approaches with pre-trained code models in various software engineering tasks. For instance, Wang et al.~\cite{sigsoft/WangYGP0L22} investigated the effectiveness of prompt learning in code intelligence tasks, such as clone detection and code summarization. Another study by Wang et al.~\cite{kbse/WangLLP23} utilized pre-trained code models as knowledge bases and employed prompt learning for variable explanation.
In this study, we strategically fine-tune pre-trained code models to facilitate practical type inference.
\section{Problem Definition}\label{sec:definition}

\begin{figure}
    \includegraphics[width=0.95\columnwidth]{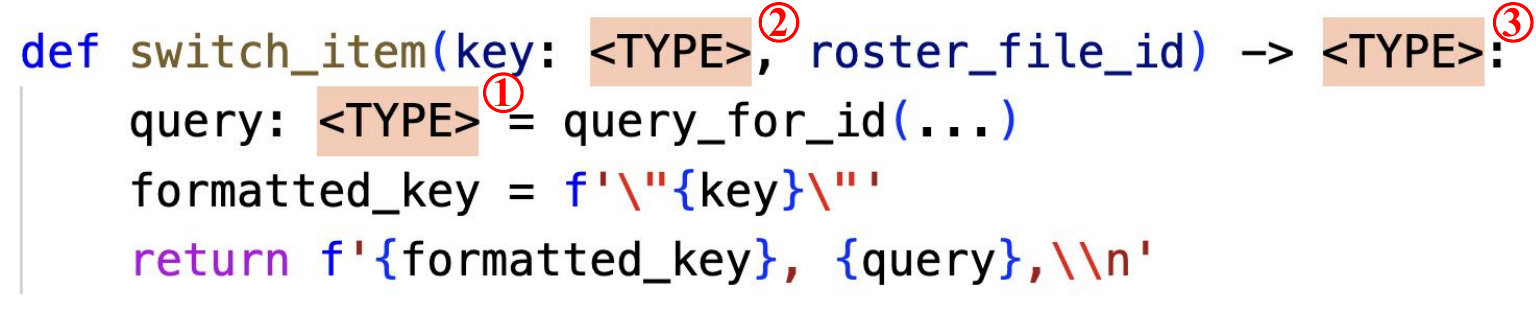}
    \caption{Type Placeholders for Three Categories of Variables}
    \label{fig:placeholders}
\end{figure}

We define the type inference problem within a Python function as a cloze-style \emph{fill-in-blank} task~\cite{emnlp/FengGTDFGS0LJZ20,iclr/GuoRLFT0ZDSFTDC21,emnlp/0034WJH21,iclr/FriedAL0WSZYZL23}, focusing on three categories of variables: local variables, function arguments, and return types. To annotate the type for a target variable, we insert a type placeholder based on its category, adhering to the annotation guidelines outlined in the PEPs. In Figure~\ref{fig:placeholders}, the three numbered ``\orangebox{\mask}'' represent the type placeholders inserted for local variables, function arguments, and return types, respectively.
In practical applications, we address one target variable in a given Python function at a time, following the approach adopted in previous studies~\cite{sigsoft/PradelGL020,icse/MirLPG22,kbse/PengWWGL23}. Functions containing type placeholders are referred to as \emph{type-missed functions}, where suitable types need to be predicted to fill the placeholders with difference categories of types, including builtin types, parameterized types, and use-defined or third-party types.


\section{Approach}\label{sec:approach}

\begin{figure*}
    \centering
    \includegraphics[width=1.8\columnwidth]{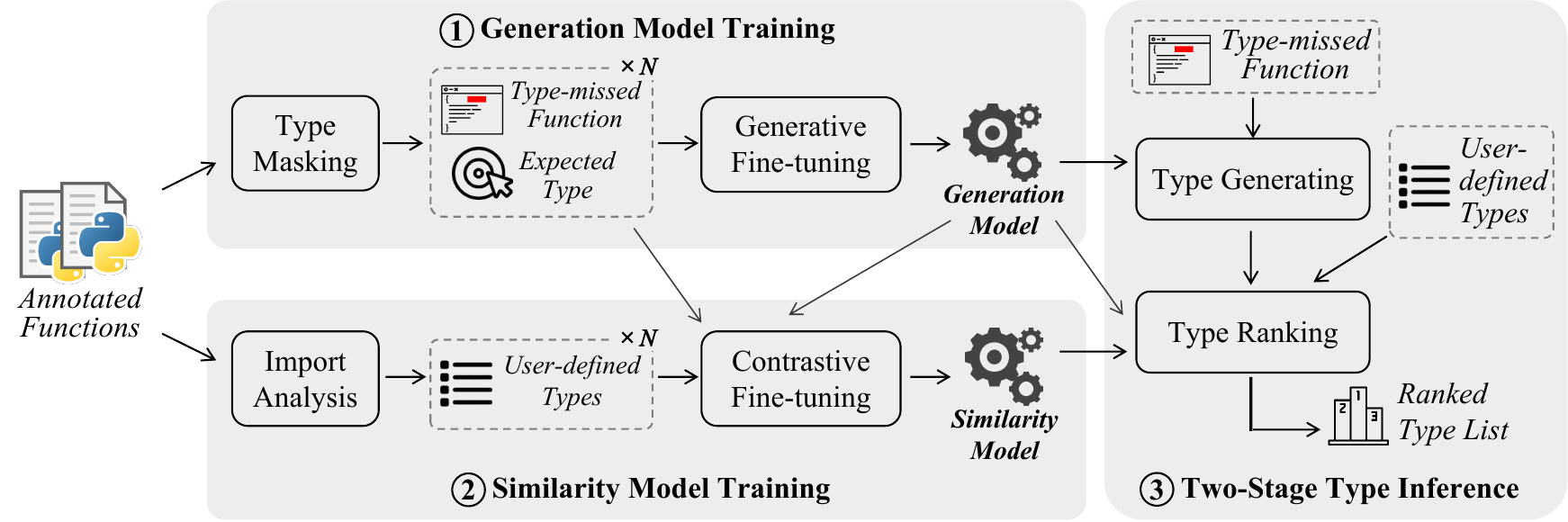}
    \caption{ Overview of \app}
    \label{fig:overview}
\end{figure*}

The overview of \app is presented in Figure~\ref{fig:overview}. \app first offline trains a generation model and a similarity model by fine-tuning existing pre-trained code models (\ie base models) using annotated Python functions (\ie \ding{172} and \ding{173}). The trained models are used to support the online two-stage \emph{generating-then-ranking} inference (\ie \ding{174}).

\ding{172} When training the generation model, \app employs a process where it masks variable types in annotated functions, replacing them with type placeholders to create type-missed functions. These type-missed functions, along with the corresponding expected types, are then utilized for fine-tuning a base model. The fine-tuning process is guided by a generative span masking objective~\cite{emnlp/0034WJH21}, which directs the model to generate appropriate types to fill the placeholders.

\ding{173} For the training of the similarity model, \app conducts an analysis of import statements to identify user-defined types available for the type-missed functions. These identified types are then combined with the types generated by the generation model to form a comprehensive set of candidate types. \app employs a contrastive training objective~\cite{icml/ChenK0H20,corr/abs-1807-03748,nips/OordVK17,cvpr/He0WXG20} to fine-tune a base model, instructing it to differentiate expected types from all candidate types based on their contextual similarities with the type-missed functions.

\ding{174} With the trained generation model and similarity model, \app executes type inference for a given type-missed function that contains a type placeholder. This process adheres to the \gtr inference framework, which encompasses both the generation of candidate types and their subsequent ranking alongside available user-defined types. In the ranking stage, candidates are ranked by combining the generative likelihood and the contextual similarity yeilded by the generation model and similarity model, respectively.

\subsection{Model Architecture Selection}
We select encoder-decoder architecture models, such as CodeT5~\cite{emnlp/0034WJH21}, as the base model for the \gtr framework and provide a concise overview of its encoding-decoding process. 

\subsubsection{Encoder-Decoder Model}
Encoder-decoder models typically undergo specific pre-training on a code corpus with the masked span prediction task, compelling the model to fill in masked span (token sequence) placeholders within the code. Moreover, with the encoder-decoder architecture, such models can capture bidirectional context within code while generating token sequences of arbitrary length and combination to fill the placeholders~\cite{kbse/WangLLP23}. This dual capability perfectly aligns with our fill-in-blank type inference task, distinguishing it from encoder-only models (\eg CodeBERT~\cite{emnlp/FengGTDFGS0LJZ20} and GraphCodeBERT~\cite{iclr/GuoRLFT0ZDSFTDC21}) and decoder-only models (\eg InCoder~\cite{iclr/FriedAL0WSZYZL23}).

\subsubsection{Encoding-Decoding Process}\label{sec:encoding-decoding}
The encoding-decoding process of an encoder-decoder model is a critical aspect for the fill-in-blank task of type inference.  

\begin{itemize}[leftmargin=15pt]
    \item \textbf{Encoding.} After tokenizing a given type-missed function into a token sequence $X=[\bos, x_1, x_2, ..., x_M, \eos]$ (where ``\bos'' and ``\eos'' denote the beginning and end of the sequence), the encoder processes $X$ and generates corresponding hidden states $\bm{H}_X = [\bm{hx}_B, \bm{hx}_1, \bm{hx}_2, ..., \bm{hx}_M, \bm{hx}_E]$ (\ie feature vectors). These vectors capture bidirectional contextual information from $X$, and the encoding process is mathematically expressed as:
    \begin{equation}\label{eq:encoder}
        \bm{H}_X \leftarrow \text{Encoder}(X)
    \end{equation}

    \item \textbf{Decoding.} The resulting $\bm{H}_X$ is then fed into the decoder, tasked with sequentially processing an input partial token sequence and predicting the next token. Specifically, at each decoding step (the $t$-th step), the decoder takes $\bm{H}_X$ and the partial sequence $Y^{:t-1} = [\bos, y_1, y_2, ..., y_{t-1}]$ as input to compute the corresponding hidden states $\bm{H}_Y^{:t-1} = [\bm{hy}_B, \bm{hy}_1, \bm{hy}_2, ..., \bm{hy}_{t-1}]$ for $Y^{:t-1}$:
    \begin{equation}\label{eq:decoder}
        \bm{H}_Y^{:t-1} \leftarrow \text{Decoder}(\bm{H}_X, Y^{:t-1})
    \end{equation}
    Subsequently, the decoder then utilizes a classification head (\ie linear layers) on $\bm{hy}_{t-1}$ to predict a $|\mathcal{V}|$-dimensional probability distribution $\bm{p}_{t}$, corresponding to the token vocabulary $\mathcal{V}$, as expressed by:
    \begin{equation}\label{eq:head}
        \bm{p}_{t} \leftarrow \text{Head}(\bm{hy}_{t-1})
    \end{equation}
    Guided by the distribution $\bm{p}$, the decoder predicts the next token $y'_t$, either by greedily selecting the token with the highest probability or employing specific sampling techniques~\cite{nips/BengioVJS15,acl/LewisDF18,iclr/HoltzmanBDFC20}. 
    
    The decoding process concludes when the ``\eos'' is fed. After finishing all the decoding steps on the input token sequence $Y = [\bos, y_1, y_2, ..., y_N]$, we denote the corresponding hidden states and predicted token sequences by the decoder as $\bm{H}_Y = [\bm{hy}_B, \bm{hy}_1, \bm{hy}_2, ..., \bm{hy}_{N}]$ and $Y' = [y'_1, y'_2, ..., y'_N, \eos]$, respectively.
\end{itemize}
\subsection{Generation Model Training}
The training of the generation model within \app commences with the retrieval of type annotations present in Python functions, adhering to the guidelines outlined in the PEP proposals. Subsequently, fine-tuning of a base model (\ie a pre-trained encoder-decoder model) is carried out, employing a generative span masking objective~\cite{emnlp/0034WJH21,jmlr/RaffelSRLNMZLL20}.

\subsubsection{Type Annotation Masking}
\begin{figure}
    \centering
    \includegraphics[width=0.85\columnwidth]{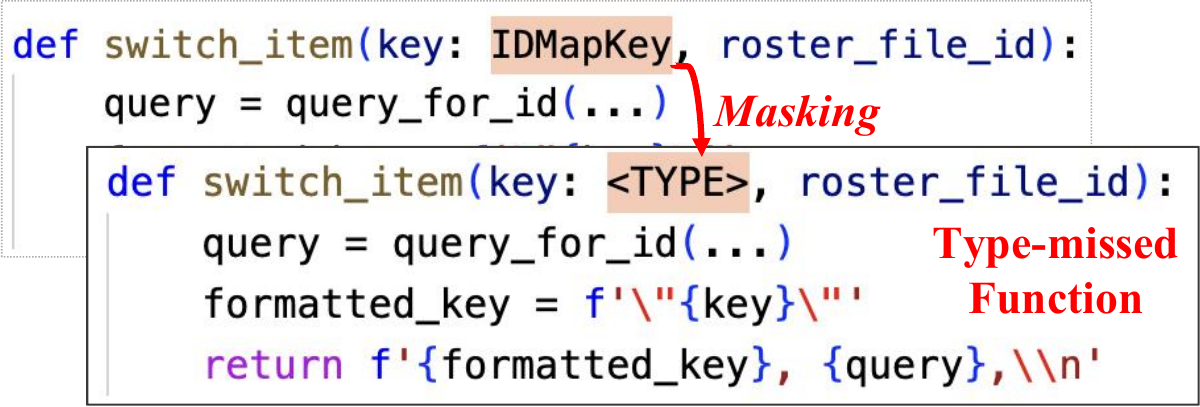}
    \caption{Type Annotation Masking with Type Placeholder}
    \label{fig:masking}
\end{figure}

Given a Python function with annotated local variables, function arguments, and return types following PEPs' guidelines, \app systematically processes these type annotations. It applies a masking procedure to each annotation, replacing them with the placeholder ``\mask''. This operation results in the creation of a type-missed function, complemented by the corresponding expected type. For instance, by masking the type annotation ``IDMapKey'' associated with the function argument ``key'', \app generates a type-missed function containing a type placeholder, as illustrated in Figure~\ref{fig:masking}, accompanied by the expected type ``IDMapKey''. After processing all annotated Python functions, \app obtains a training data that consists of type-missed functions and the corresponding expected types.

\subsubsection{Generative Fine-tuning}\label{sec:generitive-tuning}
\app proceeds with the training of the generation model, involving the fine-tuning of a base model based on acquired type-missed functions and their corresponding expected types. For each pair comprising a type-missed function $func$ and its corresponding expected type $type$, \app initiates the process by tokenizing them into their respective token sequences, denoted as $X$ and $Y$, using the model's tokenizer. Subsequently, \app employs the encoding-decoding process of the base model, as detailed in Section~\ref{sec:encoding-decoding}, and calculates the loss.

Specifically, \app employs the encoder to generate the hidden states $\bm{H}_X$ for $X$ (\cf Equation~\ref{eq:encoder}). Subsequently, in the decoding process, it feeds the first $(t-1)$ tokens in $Y$ (\ie $Y^{:t-1}$) into the decoder, computing the probability distribution $\bm{p}_{t}$ using Equations~\ref{eq:decoder} and \ref{eq:head}. Following this, the cross-entropy loss is computed based on the expected next token $y_{t}$ and $\bm{p}_{t}$, adhering to the span masking objective~\cite{emnlp/0034WJH21,jmlr/RaffelSRLNMZLL20} that is commonly utilized in the pre-training encoder-decoder models~\cite{emnlp/0034WJH21}. The fine-tuning process involves minimizing this loss, guiding the optimization of model parameters using the Adam optimizer~\cite{corr/KingmaB14}, which aims to enhance the model's capability to maximize the probability of predicting $y_{t}$ within $\bm{p}_{t}$.

\subsubsection{Resulting Generation Model}\label{sec:generation-model}
After fine-tuning, the resulting generation model in \app serves two primary purposes:
\begin{itemize}[leftmargin=15pt]
    \item \textbf{Candidate Type Generation.} When presented with a type-missed function $func$, \app initially tokenizes it into the corresponding token sequence $X$. Subsequently, \app feeds $X$ into the generation model's encoder and utilizes the decoder to generate a token sequence $Y'$, adhering to the encoding-decoding process outlined in Section~\ref{sec:encoding-decoding}. It is essential to note that, during the decoding process, the previously predicted tokens serve as the decoder's input to predict the next token. In other words, when predicting the next token $y'_{t}$ at the $t$-th decoding step, the input partial token sequence $Y^{:t-1}$ (\cf Section~\ref{sec:encoding-decoding}) comprises the previous predicted $(t-1)$ tokens (\ie $Y^{:t-1} = [\bos, y'_{1}, ..., y'_{t-1}]$) by the decoder. Upon completing the decoding process, the generated token sequence $Y'$ undergoes conversion into a candidate type. Additionally, \app integrates beam search into the decoding process, generating multiple token sequences and, consequently, yielding multiple candidate types for the placeholder within the given type-missed function. 
    \item \textbf{Generative Likelihood Computation.} When presented with a type-missed function $func$ and a candidate type $type$, \app initiates the process by tokenizing them into the corresponding token sequences $X$ and $Y$. Employing the encoding-decoding process detailed in Section~\ref{sec:encoding-decoding}, \app feeds $X$ into the generation model's encoder. Subsequently, it sequentially inputs the tokens in $Y$ into the decoder to compute the generative likelihood of $Y$. Specifically, at the $t$-th decoding step, the decoder takes as input the initial $t-1$ tokens in $Y$, \ie $Y^{:t-1}=[\bos, y_{1}, ..., y_{t-1}]$, and predicts the probability distribution $\bm{p}_t$. \app retrieves the probability of the next token $y_t$ in $Y$ from $\bm{p}_t$, denoted as $p(y_t)$, and computes the generative likelihood of $type$ by multiplying all probabilities when finishing inputting all decoding steps on $Y$:
    \begin{equation}\label{eq:likelihood}
        lik(func, type) = \prod_{y_t \in Y} p(y_t)
    \end{equation}
\end{itemize}

\subsection{Similarly Model Training}
The purpose of the similarity model is to effectively gauge the contextual similarity between the given type-missed function and the candidate types. To achieve this, \app employs a contrastive training objective~\cite{icml/ChenK0H20,corr/abs-1807-03748,nips/OordVK17,cvpr/He0WXG20}, which seeks to cultivate generalizable and robust representations for both type-missed functions and candidate types. This objective achieves this goal by maximizing the similarity between the type-missed function and the expected type while minimizing the similarity between the type-missed function and other candidate types.

\subsubsection{Import Analysis}\label{sec:import}
\app incorporates user-defined types for the type placeholder within the given type-missed function while constructing the training dataset for contrastive learning. This data is derived by analyzing import statements existing in the same project environment as the type-missed function, in accordance with established practices~\cite{icse/Ye0DW022,kbse/PengWWGL23}. Specifically, \app parses the current source file containing the target function and gathers candidate user-defined types from two distinct sources.
Firstly, all type definitions in the current file are directly considered as candidate user-defined types. Additionally, \app examines import statements in the current file to identify the source files being imported. It then includes the types defined in these imported files in the list of candidate user-defined types. The gathered user-defined types might be project-specific, with names that are not encountered (unseen) in other projects.

\subsubsection{Contrastive Fine-tuning}\label{sec:contrastive-tuning}
The fine-tuning process for the similarity model consists of three pivotal components: negative instance construction, similarity computation, and model training. Adhering to the terminology commonly used in contrastive training conventions, we designate the type-missed function as the \emph{anchor}, the corresponding expected type as the \emph{positive instance}, and other candidates as \emph{negative instances}.

\begin{itemize}[leftmargin=15pt]
    \item \textbf{Negative Instance Construction.} 
    Constructing suitable negative instances for the expected type is crucial for effective contrastive training. \app achieves this by generating $K$ negative instances from two distinct sources. For the given type-missed function (\ie anchor), the first source utilizes the trained generation model, employing a beam search algorithm during model decoding to generate $K$ candidates (\cf Section~\ref{sec:generation-model}). These candidates typically encompass builtin types, commonly used third-party types, and parameterized types. The second source involves user-defined types obtained through import analysis. \app combines the types from both sources, excluding the positive instance, and randomly selects $K$ ones as the negative instances.
    
    \item \textbf{Similarity Computation.} 
    To quantify contextual similarity between the anchor, namely $func$, and a positive/negative instance, namely $type$, \app tokenizes them into their respective token sequences, $X$ and $Y$. Subsequently, these sequences undergo processing by a pre-trained base model (\ie a pre-trained encoder-decoder model), executing the encoding-decoding process described in Section~\ref{sec:encoding-decoding}.
    Upon completion of the process, \app computes the average of the hidden states $\bm{H}_X$ produced by the encoder as the vector representation of $func$, and the average of the hidden states $\bm{H}_Y$ produced by the decoder as the vector representation of $type$.
    The two representations, namely $\text{AVG}(\bm{H_X})$ and $\text{AVG}(\bm{H_Y})$, are employed to assess the contextual similarity between $func$ and $type$ by calculating their cosine similarity:
    \begin{equation}\label{eq:similarity}
        \text{sim}(func, type) = \text{cosine}(\text{AVG}(\bm{H}_X), \text{AVG}(\bm{H}_Y))
    \end{equation}
    
    \item \textbf{Model Training.} 
    Given the anchor $func$, \app computes the InfoNCE loss~\cite{cvpr/He0WXG20,corr/abs-1807-03748} based on it, its corresponding positive instance $pos$, and the generated negative instances $\mathbb{N}$. The loss function is defined as follows:
    \begin{equation}
        \mathcal{L} = - \text{log} \frac{\text{e}^{\text{sim}(func, pos)}}{\text{e}^{\text{sim}(func, pos)} + \underset{neg \in \mathbb{N}}\sum \text{e}^{\text{sim}(func, neg)}} \nonumber
    \end{equation}
    Maximizing this loss function guides the model to enhance the disparity between the similarities of the positive instance and the negative instances, allowing the model to learn to distinguish the most suitable type for the given type-missed function. The optimization process utilizes the Adam optimizer~\cite{corr/KingmaB14} to update the model's parameters.
\end{itemize}

\subsubsection{Resulting Similarity Model}\label{sec:similarity-model}
Following fine-tuning, the resulting similarity model demonstrates an effective capability to measure the contextual similarity between the provided type-missed function and its associated candidate types, as depicted in Equation~\ref{eq:similarity}. Due to the contrastive training objective and the incorporation of negative instances from two distinct sources, the similarity model exhibits strong generalizability and robustness~\cite{corr/abs-1807-03748}, encompassing both generated candidates and user-defined types.

\subsection{Two-Stage Type Inference}\label{sec:inference}
Leveraging both the generation model and the similarity model, \app employs the two-stage \gtr strategy for type inference, as depicted in Figure~\ref{fig:framework}.

\subsubsection{Type Generating}
When presented with a function and a variable slated for annotation, \app initially inserts a type placeholder after the variable based on its category (\ie local variable, function argument, or return value), creating a type-missed function $func$ akin to the example in Figure~\ref{fig:masking}. Subsequently, this type-missed function undergoes processing by the generation model, employing beam search during decoding to generate $K$ token sequences, thereby forming $K$ candidate types (\cf Section~\ref{sec:generation-model}).

\subsubsection{Type Ranking}
\app conducts import analysis, as detailed in Section~\ref{sec:import}, to acquire user-defined types available for the type placeholder in $func$. These user-defined types, combined with the generated candidates, constitute a more comprehensive candidate pool. Note that if a generated candidate is neither a built-in type nor a parameterized type whose base is a built-in type, and it is not found in the user or library-defined types, we exclude it from the candidate pool. Subsequently, \app ranks all candidates in the pool by considering both generative likelihood and contextual similarity for each candidate. For a given candidate $cand$ in the pool, \app computes its generative likelihood using the generation model (\cf Section~\ref{sec:generation-model}) and its contextual similarity with $func$ using the similarity model (\cf Section~\ref{sec:similarity-model}).

The final score of candidate $cand$ is determined by the sum of generative likelihood $\text{lik}(func, cand)$ and contextual similarity $\text{sim}(func, cand)$, expressed as:
\begin{equation}\label{eq:final-score}
    \text{score}(cand) = \text{lik}(func, cand) + \text{sim}(func, cand)
\end{equation}
Ultimately, all candidates for the given type-missed function are ranked based on their scores in descending order, forming a ranked list of types for the variable intended for annotation.

The inclusion of both generative likelihood and contextual similarity in the score calculation is motivated by specific considerations, inspired by the methodology utilized in contrastive text generation~\cite{nips/AnFLKQ022,nips/SuLWYKC22}. Solely relying on generative likelihood, as discussed in Section~\ref{sec:generation-model}, could lead to biased rankings. During training, the decoding process is guided by the ground truth input, which is absent during generation~\cite{acl/AroraABC22}. This mismatch introduces exposure bias, making the decoding process less robust in distinguishing between ``good'' and ``bad'' tokens in the certain decoding steps.
In contrast, the similarity model generates robust and generalizable representations for both the generated candidates and the user-defined types. This is achieved by combining these two sources during the construction of negative instances for contrastive learning~\cite{corr/abs-1807-03748}. Therefore, by integrating both generative likelihood and contextual similarity, \app capitalizes on the task alignment benefits of the generation model and the generalizability and robustness of the similarity model. This dual consideration enhances the overall effectiveness of the \gtr inference process.

\section{Evaluation}\label{sec:evaluation}

We conduct extensive experiments to assess the effectiveness of \app in Python type inference from various dimensions. 
The research questions are summarized as follows:

\begin{itemize}
    \item \textbf{RQ1 (Effectiveness)}: How effectively can \app infer types for variables? How effectively can \app handle with various categories of types, especially parameterized types and (unseen) user defined types?
    \item \textbf{RQ2 (Robustness)}: How robust is \app in inferring types for different categories of variables, including local variables, function arguments, and return types?
    \item \textbf{RQ3 (Efficiency)}: How effectively can \app infer types for different sources of types, encompassing elementary types, parameterized types, and (unseen) user-defined types?
    \item \textbf{RQ4 (Ablation Study)}: Are the two stages in \gtr useful for type inference?
\end{itemize}

\subsection{Setup}
We present the experimental setup for the evaluation, encompassing the implementation, dataset, baselines, and metrics.

\subsubsection{Implementation}
In our current implementation, we utilize CodeT5 as the base model for training both the generation and similarity models, given its widespread use as an encoder-decoder pre-trained model for code. We implement \app in Python using the Transformers library~\cite{Transformers}, a widely-used library for language models. The base models for training the generation and similarity models is downloaded from the Hugging Face Hub~\cite{codet5base}. Following established practices~\cite{icse/MirLPG22,icse/PengGLG0ZL22,kbse/PengWWGL23}, we consider the top-5 candidate types for predictions, setting $K=5$ (\ie beam size) during both the training and inference stages of \app. The training hyperparameters for both the generation model and similarity model include a training epoch of 3, a learning rate of 1e-5, and a training batch size of 8. 

\subsubsection{Dataset}
In line with prior studies~\cite{icse/MirLPG22,icse/PengGLG0ZL22}, we conduct the evaluation on the ManyTypes4Py dataset~\cite{msr/MirLG21}, which is partitioned into training and test sets with an 8:2 project ratio (no overlapping projects between training set and test set), resulting in 242,954 instances for training and 85,205 instances for testing. For the \typegen evaluation, the authors further sampled 10,000 instances from the entire test set to reduce the computational cost associated with calling large language models (LLMs). To maintain consistency in our experimental setup, we employ this sampled subset~\cite{dataset} as the definitive test set for all experiments pertaining to RQ1-RQ4. Among the test instances involving user-defined types, 579 instances require the inference of types that are \textbf{previously unseen in training set}. These unseen types are challenging to predict and provide insights into the practicality of the inference approach. The training set is utilized for training both \app and baselines, as well as providing few-shot examples for the prompt construction in \typegen. The test set is then employed to assess the type inference effectiveness of \app and baselines. Table~\ref{tab:datasets} provides an overview of the training and test sets, illustrating the distributions of different type and variable categories. ``Ele'', ``Par'', and ``Usr (Unseen)'' represent different type categories~\cite{kbse/PengWWGL23}, denoting Elementary, Parameterized, and (unseen) user-defined Types, respectively. ``Var'', ``Arg'', and ``Ret'' represent different variable categories~\cite{kbse/PengWWGL23}, denoting Local Variables, Function Arguments, and Return Values, respectively. 

\begin{table}
    \caption{Overview of training and test sets.}
    \label{tab:datasets}
    \centering
    \renewcommand{\arraystretch}{1.5}
    \setlength{\tabcolsep}{2.5pt}
    \begin{tabular}{ccccccccc}
        \Xhline{2\arrayrulewidth}
        \multirow{2}{*}{\textbf{Dataset}}  &  & \multicolumn{3}{c}{\textbf{Type Category}}                                                               &  & \multicolumn{3}{c}{\textbf{Variable Category}}                                                            \\ \cline{3-5} \cline{7-9} 
                                           &  & \textbf{\#Ele} & \textbf{\#Par} & \textbf{\#Usr (\#Unseen)} &  & \textbf{\#Var} & \textbf{\#Arg} & \textbf{\#Ret} \\ 
        \Xhline{1.5\arrayrulewidth}
        \textbf{Training} &  & 128,006                         & 67,185                          & 47,763 (-)                          &  & 172,459                         & 48,461                          & 22,034                         \\
        (242,954)                          &  & 52.7\%                          & 27.6\%                          & 19.7\% (-)                         &  & 70.9\%                          & 20.0\%                          & 9.1\%                           \\ \hline
        \textbf{Test}     &  & 5,199                           & 2,748                           & 2,053 (579)                          &  & 7,091                           & 1,995                           & 914                             \\
        (10,000)                           &  & 52.0\%                          & 27.5\%                          & 20.5\% (5.8\%)                         &  & 70.9\%                          & 20.0\%                          & 9.1\%                           \\ 
        \Xhline{2\arrayrulewidth}
    \end{tabular}
\end{table}

\subsubsection{Baselines}
We include several state-of-the-art learning-based type inference approaches as baselines. We omit rule-based approaches from consideration, given that learning-based methods have demonstrated superior effectiveness in prior studies~\cite{icse/PengGLG0ZL22}.

\begin{itemize}
    
    \item \textbf{\typewriter}~\cite{sigsoft/PradelGL020}: A classification-based type inference approach that utilizes RNNs to encode various code features (\eg identifiers and code tokens) to predict types for target variables. Note that \typewriter cannot process local variables; therefore, the metrics are calculated only on the subset consisting of function arguments and return types.
    \item \textbf{\typeforpy}~\cite{icse/MirLPG22}: A similarity-based type inference approach classifies new Python programs by assigning them to a specific type clusters, using those clusters as the predicted types for the target variables.
    \item \textbf{HiTyper}~\cite{icse/PengGLG0ZL22}: A hybrid type inference approach combining static inference and similarity-based models by leveraging type dependency graphs (TDGs) to record and integrate type dependencies among variables.
    \item \textbf{CodeT5-\emph{zs}}: A pre-trained CodeT5-base model~\cite{emnlp/FengGTDFGS0LJZ20} utilized for zero-shot variable type inference, treating the problem as a cloze-style fill-in-the-blank task, similar to \app.
    \item \textbf{CodeT5-\emph{ft}}: A fine-tuned CodeT5-base model~\cite{emnlp/FengGTDFGS0LJZ20} trained with the same training data of \app, also treating the problem as a cloze-style fill-in-the-blank task.
    \item \textbf{\typegen}~\cite{kbse/PengWWGL23}: An LLM-generation-based type inference approach that leverages prompt engineering techniques, including in-context learning and chain-of-thoughts, to harness the capability of LLM for generating types for target variables.

\end{itemize}


For the baselines, we utilize the released replication implementations and data. We opt not to compare our approach with the other two learning-based type inference methods, namely the classification-based \textbf{DLInfer}~\cite{icse/YanFFX23} and the generation-based \textbf{TypeT5}~\cite{iclr/WeiDD23}. These approaches heavily rely on static analysis and were originally trained and evaluated on smaller datasets (700 and 663 projects, respectively). Despite our attempt to apply DLInfer and TypeT5 to the larger ManyTypes4Py dataset (5,996 projects), scalability issues rendered them inapplicable. Additionally, methodological differences, such as DLInfer segmenting source code into syntax-broken short snippets (\eg incomplete statements) and TypeT5 not processing local variables within functions, make a fair comparison challenging.

\begin{table*}
    \caption{Inference accuracy of \app and baseline approaches. CLS, SIM, and GEN represent Classification-based, Similarity-base, and Generation-based, respectively. The \first{best} and \second{second-best} results for each category are are highlighted.}\label{tab:overall}
    \centering
    \renewcommand{\arraystretch}{1.5}
    \setlength{\tabcolsep}{3pt}
    \begin{threeparttable}
        \begin{tabular}{cccccccccccccccccccc} 
            \Xhline{2\arrayrulewidth}
            \multirow{2}{*}{\textbf{Metric}}                                                         &  & \multirow{2}{*}{\textbf{Approach}} &  & \multirow{2}{*}{\textbf{Catetory}} &  & \multicolumn{4}{c}{\textbf{Top-1 (\%)}}                                                     &  & \multicolumn{4}{c}{\textbf{Top-3 (\%)}}                                                &  & \multicolumn{4}{c}{\textbf{Top-5 (\%)}}                                                    \\ 
            \cline{7-10}\cline{12-15}\cline{17-20}
                                                                                                     &  &                                    &  &                                    &  & \textbf{Ele}      & \textbf{Par}      & \textbf{Usr (Unseen)}              & \textbf{All}   &  & \textbf{Ele}   & \textbf{Par}      & \textbf{Usr~(Unseen)}            & \textbf{All}   &  & \textbf{Ele}      & \textbf{Par}      & \textbf{Usr~(Unseen)}            & \textbf{All}    \\ 
            \Xhline{1.5\arrayrulewidth}
            \multirow{8}{*}{\begin{tabular}[c]{@{}c@{}}\textbf{Exact}\\ \textbf{Match} \\ \textbf{(EM)} \end{tabular}} &  & TypeWriter\tnote{*}                        &  & CLS                                &  & \other{69.9}      & \other{43.0}      & \other{36.5} (\other{30.0})        & \other{55.0}   &  & \other{78.8}   & \other{53.4}      & \other{39.3} (\other{32.6})      & \other{62.4}   &  & \other{84.5}      & \other{59.7}      & \other{42.4} (\other{36.5})      & \other{67.3}    \\ 
            \cline{3-20}
                                                                                                     &  & Type4Py                            &  & \multirow{2}{*}{SIM}               &  & \first{95.1}      & \other{67.9}      & \other{43.0} (\,\other{6.9}\,)\,\, & \other{76.9}   &  & \other{95.3}   & \other{71.7}      & \other{49.5} (\other{10.0})      & \other{79.5}   &  & \other{95.4}      & \other{73.5}      & \other{50.9} (\other{10.9})      & \other{80.3}    \\
                                                                                                     &  & HiTyper\tnote{*}                            &  &                                    &  & \other{94.0}      & \other{60.0}      & \other{61.3} (\other{51.3})        & \other{76.3}   &  & \other{94.2}   & \other{60.4}      & \other{68.0} (\other{56.3})      & \other{78.7}   &  & \other{94.3}      & \other{60.7}      & \other{71.6} (\other{57.5})      & \other{79.6}    \\ 
            \cline{3-20}
                                                                                                     &  & CodeT5-\textit{zs}                 &  & \multirow{3}{*}{GEN}               &  & \other{34.1}      & \other{15.9}      & \other{35.3} (\other{33.2})      & \other{29.4}   &  & \other{33.0}   & \other{18.9}      & \other{36.5} (\other{37.7})      & \other{29.9}   &  & \other{33.5}      & \other{21.2}      & \other{37.7} (\other{38.7})      & \other{31.0}    \\
                                                                                                     &  & CodeT5-\textit{ft}                 &  &                                    &  & \other{94.0}      & \second{73.9}     & \other{73.2} (\other{61.3})        & \second{84.2}  &  & \second{97.0}  & \second{86.7}      & \other{78.8} (\other{66.3})      & \second{90.4}  &  & \first{97.7}      & \first{89.4}      & \other{80.8} (\other{67.7})      & \second{91.9}   \\
                                                                                                     &  & TypeGen                            &  &                                    &  & \other{89.8}      & \other{60.0}      & \second{79.0} (\second{63.6})       & \other{79.2}   &  & \other{93.8}   & \other{74.6}      & \second{83.7} (\second{69.9})    & \other{86.2}   &  & \other{94.3}      & \other{77.8}      & \second{84.6} (\second{70.8})    & \other{87.5}    \\ 
            \cline{3-20}
                                                                                                     &  & \app                               &  & GEN+SIM                            &  & \second{94.5}     & \first{75.1}      & \first{80.2} (\first{67.5})       & \first{86.2}   &  & \first{97.5}   & \first{87.2}     & \first{90.2} (\first{79.8})      & \first{93.2}   &  & \first{97.7}     & \first{89.4}     & \first{94.1} (\first{85.0})      & \first{94.6}    \\
                                                                                                     \rowcolor{gray!20}
                                                                                \cellcolor{white}    & \cellcolor{white}  & \multicolumn{3}{c}{$\Delta$ to the best baseline}        &  & $\downarrow$0.6 & $\uparrow$1.6    & $\uparrow$1.5~~($\uparrow$6.1) & $\uparrow$2.4 &  & $\uparrow$0.5 & $\uparrow$0.6 & $\uparrow$7.8 ($\uparrow$14.2)  & $\uparrow$3.1 &  & 0.0 & 0.0 & $\uparrow$11.2 ($\uparrow$20.1) & $\uparrow$2.9  \\ 
            \Xhline{1.5\arrayrulewidth}
            \multirow{8}{*}{\begin{tabular}[c]{@{}c@{}}\textbf{Base}\\ \textbf{Match} \\ \textbf{(BM)}\end{tabular}}  &  & TypeWriter\tnote{*}                         &  & CLS                                &  & \other{71.9}      & \other{49.6}      & \other{37.3} (\other{31.8})        & \other{57.5}   &  & \other{82.1}   & \other{68.6}      & \other{41.2} (\other{36.5})      & \other{67.4}   &  & \other{88.3}      & \other{79.5}      & \other{45.8} (\other{42.5})      & \other{73.9}    \\ 
            \cline{3-20}
                                                                                                     &  & Type4Py                            &  & \multirow{2}{*}{SIM}               &  & \other{95.3}      & \other{79.2}      & \other{45.3} (\other{10.2})        & \other{80.6}   &  & \other{95.7}   & \other{85.4}      & \other{53.8} (\other{17.8})      & \other{84.3}   &  & \other{95.7}      & \other{87.9}      & \other{56.6} (\other{21.4})      & \other{85.5}    \\
                                                                                                     &  & HiTyper\tnote{*}                            &  &                                    &  & \other{94.4}      & \other{84.5}      & \other{61.3} (\other{51.3})        & \other{85.1}   &  & \other{94.6}   & \other{88.2}      & \other{68.0} (\other{56.3})      & \other{87.6}   &  & \other{94.8}      & \other{90.4}      & \other{71.6} (\other{57.5})      & \other{89.1}    \\ 
            \cline{3-20}
                                                                                                     &  & CodeT5-\textit{zs}                 &  & \multirow{3}{*}{GEN}               &  & \other{38.2}      & \other{53.4}      & \other{36.8} (\other{36.8})       & \other{42.1}   &  & \other{37.7}   & \other{57.3}      & \other{39.4} (\other{42.1})      & \other{43.4}   &  & \other{39.1}      & \other{60.5}      & \other{42.2} (\other{44.7})      & \other{45.6}    \\
                                                                                                     &  & CodeT5-\textit{ft}                 &  &                                    &  & \second{95.5}     & \first{93.3}      & \other{73.8} (\other{63.0})       & \second{90.5}  &  & \second{97.5}  & \first{96.9}      & \other{79.4} (\other{67.9})      & \second{93.7}  &  & \first{98.0}      & \second{97.8}      & \other{81.5} (\other{69.6})      & \second{94.6}   \\
                                                                                                     &  & TypeGen                            &  &                                    &  & \other{91.3}      & \other{87.1}      & \second{79.6} (\second{63.9})      & \other{87.4}   &  & \other{94.7}   & \other{91.8}      & \second{84.4} (\second{70.6})    & \other{91.5}   &  & \other{95.1}      & \other{92.6}      & \second{85.1} (\second{71.5})    & \other{92.0}    \\ 
            \cline{3-20}
                                                                                                     &  & \app                               &  & GEN+SIM                            &  & \first{96.2}      & \second{93.2}     & \first{81.3} (\first{70.5})        & \first{92.3}   &  & \first{97.9}   & \second{96.8}     & \first{91.0} (\first{82.4})      & \first{96.2}   &  & \first{98.0}     & \first{97.9}     & \first{95.3} (\first{88.4})      & \first{97.4}    \\
                                                                                                     \rowcolor{gray!20}
                                                                                 \cellcolor{white}   & \cellcolor{white} & \multicolumn{3}{c}{$\Delta$ to the best baseline}        &  & $\uparrow$0.7    & $\downarrow$0.1 &   $\uparrow$2.1 ($\uparrow$10.3)    & $\uparrow$2.0       &  & $\uparrow$0.4 & $\downarrow$0.1 & $\uparrow$7.8 ($\uparrow$16.7) & $\uparrow$2.7 &  & 0.0               & $\uparrow$0.1    & $\uparrow$12.0 ($\uparrow$23.6)  & $\uparrow$3.0  \\
                                                                                                     \Xhline{2\arrayrulewidth}
        \end{tabular}
        \begin{tablenotes}
            \item[*] \footnotesize{TypeWriter and HiTyper are capable of inferring types for 2,909 and 1,032 test instances, respectively. Consequently, their metrics are calculated using only these instances.}
        \end{tablenotes}
    \end{threeparttable}
\end{table*}

\subsubsection{Metrics}
Following established practices~\cite{icse/MirLPG22,icse/PengGLG0ZL22,kbse/PengWWGL23}, we employ two commonly used metrics to evaluate the inference accuracy of \app and the baselines:
\begin{itemize}
    \item \textbf{Exact Match (EM)}: This metric calculates the ratio of type predictions made by an approach that \emph{exactly} matches the type annotated by developers.
    \item \textbf{Base Match (BM)}: This metric calculates the ratio of type predictions made by an approach that shares the common base type (\ie the outermost type) with the type annotations provided by developers. For example, ``Union[str,list]'' is base-matched, but not exact-matched, with ``Union[str,int]''.
\end{itemize}


\subsection{Effectiveness (RQ1)}
We evaluate the effectiveness of \app and the six baseline approaches in Python type inference. The detailed inference accuracy of \app and the baselines on the test set consisting of 10,000 instances is provided in Table~\ref{tab:overall}.

\subsubsection{Overall Results}
Across the entire test set (Column ``All''), our approach, \app, demonstrates Top-1, 3, and 5 Exact Match metrics of 86.2\%, 93.2\%, and 94.6\%, respectively, and Top-1, 3, and 5 Base Match metrics of 92.3\%, 96.2\%, and 97.4\%, respectively. Considering fine-grained type categories (Columns ``Ele'', ``Par'', and ``Usr (Unseen)''), \app achieves either the best or second-best results in terms of Exact Match and Base Match. Particularly noteworthy are the substantial improvements (see Rows ``$\Delta$ compared with the best baseline'') in almost all type categories, especially in unseen user-defined types (up by 20.1\% and 23.4\% in Top-5 Exact and Base Match, respectively), with minimal declines (less than 1\%) observed in few type categories such as Top-1 Exact Match for elementary types (Ele).

An interesting observation is that the Base Match metrics of all the approaches are notably higher than the Exact Match metrics for parameterized types (\ie Column ``Par'') compared with the other two type categories Columns ``Ele'' and ``Usr (Unseen)''). This discrepancy arises because inference approaches correctly predict the base type \inlinecode{Base} but face challenges predicting the specific combinations of \inlinecode{[T1,T2,...]}. 

\subsubsection{Comparison with Classification-based Approach}
\app significantly outperforms the classification-based baseline approach, TypeWriter, for all type categories in terms of Exact Match and Base Match metrics. Specifically, in Top-5 Exact Match, \app demonstrates a significant improvement over TypeWriter, with an overall enhancements of 40.6\% across the entire test set and ranging from 15.6\% to 132.9\% for fine-grained type categories. 

The superiority is attributed to the limitations of classification-based approaches, such as TypeWriter, which rely on fixed-sized type vocabularies. Such vocabularies often fail to comprehensively cover diverse parameterized types and user-defined types in their visible candidate set. \app addresses this limitation by leveraging the capacity of its generating stage to produce complex parameterized types and the ranking capability of its ranking stage to handle (unseen) user-defined types. Notably, while TypeWriter achieves about 30\%-40\% accuracy for unseen types, as its 1,000-type vocabulary covers a portion of unseen types in the ManyType4Py test set, it is unable to predict types beyond its vocabulary limit.

\subsubsection{Comparison with Similarity-based Approaches}
While demonstrating comparable accuracy for elementary types, \app significantly outperforms the similarity-based baseline approaches, Type4Py and HiTyper, for parameterized types and (unseen) user-defined types in terms of Exact Match and Base Match metrics. Specifically, in Top-5 Exact Match, \app achieves notable improvements over Type4Py and HiTyper, with enhancements of 17.8\% and 18.8\% respectively across the entire test set, ranging from 21.6\% to 679.8\% over Type4Py and from 31.4\% to 47.9\% over HiTyper for parameterized types and user-defined types.

The enhancements achieved by \app can be attributed to the following reasons. While similarity-based approaches can handle commonly used parameterized types (\eg \inlinecode{list[str]}), they struggle with inferring diverse parameterized types due to the vast number of possible combinations. In contrast, \app leverages the creative type generation capacity of its generation model during the generating stage to overcome this limitation. Moreover, in the case of (unseen) user-defined types, \app's superior performance is primarily attributed to its robustness and generalizability in the ranking stage, facilitated by the contrastive training objective~\cite{corr/abs-1807-03748}.

\subsubsection{Comparison with Generation-based Approaches}
Among the three generation-based baseline approaches, CodeT5-\emph{zs} exhibits poor accuracy due to its application in a zero-shot setting, while the fine-tuned CodeT5-\emph{ft} and TypeGen achieve the second-best or the co-best results for specific type categories and metrics. Specifically, CodeT5-\emph{ft} demonstrates effectiveness for elementary types (Column ``Ele'') and parameterized types (Column ``Par''), but is relatively ineffective for user-defined types (Column ``Usr (Unseen)''); whereas TypeGen achieves good effectiveness for user-defined types but struggles with handling parameterized types effectively. 

CodeT5-\emph{ft}'s lower efficacy in handling user-defined types stems from its inability to grasp project-specific contexts, while the limited effectiveness of TypeGen for parameterized and unseen user-defined types can be attributed to its reliance on prompt engineering techniques to leverage the capability of LLMs, which often poses challenges in designing an optimal prompt template~\cite{chi/Zamfirescu-Pereira23}. In comparison, our approach, \app, consistently achieves substantial inference accuracies for all type categories, especially (unseen) user-defined types (Column ``Usr (Unseen)''), with enhancements of 16.5\% (25.6\%) and 11.2\% (20.1\%) compared to CodeT5-\emph{ft} and TypeGen, respectively, in Top-5 Exact Match. These improvements are attributed to the comprehensive visible candidate pool and robust inference capacity integrated within the two-stage inference framework. Considering the prevalence of encountering (unseen) user-defined types in real-world projects, our approach proves to be more practical and applicable.

\summary{
    \app outperforms all six baseline approaches across the entire test set, achieving a Top-5 Exact Match metric of 93.1\%. For fine-grained type categories, \app consistently achieves (almost) the highest accuracies, particularly for unseen user-defined types, with enhancements of 25.6\% and 20.1\% compared to the best two baselines, CodeT5-\emph{ft} and TypeGen, respectively, in Top-5 Exact Match.
}

\subsection{Robustness (RQ2)}
As noted in prior research~\cite{kbse/PengWWGL23}, type inference for certain variable categories such as function arguments and return values poses greater difficulty. Hence, we delve deeper into the robustness of \app and the best two baselines in RQ1, CodeT5-\emph{ft} and TypeGen, across various variable categories. The evaluation results are outlined in Table~\ref{tab:variables}.

\begin{table}
    \caption{Top-1,3,5 Exact Match of CodeT5-\emph{ft}, TypeGen, and \app for different variable categories. The \first{best} and \second{second-best} result for each category are are highlighted.}
    \label{tab:variables}
    \centering
    \renewcommand{\arraystretch}{1.5}
    \setlength{\tabcolsep}{2.5pt}
    \begin{tabular}{ccccccccccccc} 
        \Xhline{2\arrayrulewidth}
        \multirow{2}{*}{\textbf{Approach}} & \multirow{2}{*}{} & \multicolumn{3}{c}{\textbf{Top-1 EM (\%)}}    & \multirow{2}{*}{} & \multicolumn{3}{c}{\textbf{\textbf{Top-3 EM (\%)}}} & \multirow{2}{*}{} & \multicolumn{3}{c}{\textbf{\textbf{\textbf{\textbf{Top-5 EM (\%)}}}}}  \\ 
        \cline{3-5}\cline{7-9}\cline{11-13}
                                           &                   & \textbf{Var} & \textbf{Arg} & \textbf{Ret} &                   & \textbf{Var} & \textbf{Arg} & \textbf{Ret}       &                   & \textbf{Var} & \textbf{Arg} & \textbf{Ret}                          \\ 
        \Xhline{1.5\arrayrulewidth}
        CodeT5-\textit{ft}                 &                   & \second{89.8}& 67.8         & \second{76.3}&                   & \second{94.3}& 79.4         & \second{84.4}      &                   & \second{95.3}& 82.7         & \second{85.8}                         \\
        TypeGen                            &                   & 82.0         & \second{73.9}& 69.1         &                   & 88.6         & \second{81.6}& 77.6               &                   & 90.0         & \second{83.5}& 79.4                                  \\ 
        \Xhline{1.5\arrayrulewidth}
        \app                               &                   & \first{90.6} & \first{74.6} & \first{77.7} &                   & \first{95.6} & \first{87.5} & \first{87.0}       &                   & \first{96.2} & \first{91.6} & \first{89.1}                          \\
        \Xhline{2\arrayrulewidth}
    \end{tabular}
\end{table}
Overall, \app demonstrates superior robustness across different variable categories. Specifically, CodeT5-\emph{ft} exhibits significant disparities ($\downarrow$9.1\% in Top-1 Exact Match) with \app for function arguments (Column ``Arg''), while achieving comparable results for local variables (Column ``Var'') and return values (Column ``Ret''). Conversely, TypeGen displays notable differences ($\downarrow$9.5\% and $\downarrow$11.1\% in Top-1 Exact Match) with \app for local variables (Column ``Var'') and return values (Column ``Ret''), while attaining comparable results for function arguments (Column ``Arg''). In summary, our approach, \app, consistently achieves consistently accurate results for all the three variable categories.

\app achieves consistently strong results across all three variable categories for two main reasons. First, compared to TypeGen, \app formulates the inference problem as a unified cloze-style fill-in-the-blank task, aligning perfectly with the PEP guidelines and the pre-training tasks of the base models. This allows for effective leveraging of the capacity of the pre-trained models. In contrast, TypeGen depends on distinct prompt templates tailored for each variable category, constraining the consistent utilization of LLMs across all categories. Second, compared to CodeT5-\emph{ft}, which also utilizes the unified fill-in-the-blank task formulation, the contrastive training employed in \app's similarity model training enhances robustness in discriminating multiple candidates~\cite{corr/abs-1807-03748}. Although inferring types for function arguments is particularly challenging as their corresponding definition statements are often unavailable within the given function~\cite{kbse/PengWWGL23}, \app can effectively discriminate candidate types for function arguments based on limited code context, such as the usage information of the arguments.

\summary{
In comparison to CodeT5-\emph{ft} and TypeGen, our approach, \app, showcases strong and more consistent inference accuracy across all three variable categories, particularly for function arguments and return types.
}
\subsection{Efficiency (RQ3)}
We evaluate the inference efficiency of \app alongside the best two baselines, CodeT5-\emph{ft} and TypeGen, using a set of 50 test instances randomly sampled from the test set. 
All three approaches are executed on the same machine with a stable network connection. For CodeT5-\emph{ft} and \app, a single NVIDIA V100-32G is utilized, and the test instances are processed one by one sequentially (\ie batch size of 1). For TypeGen, we adopt the default settings from their original implementation.

\begin{figure}
    \centering
    \includegraphics[width=1.0\columnwidth]{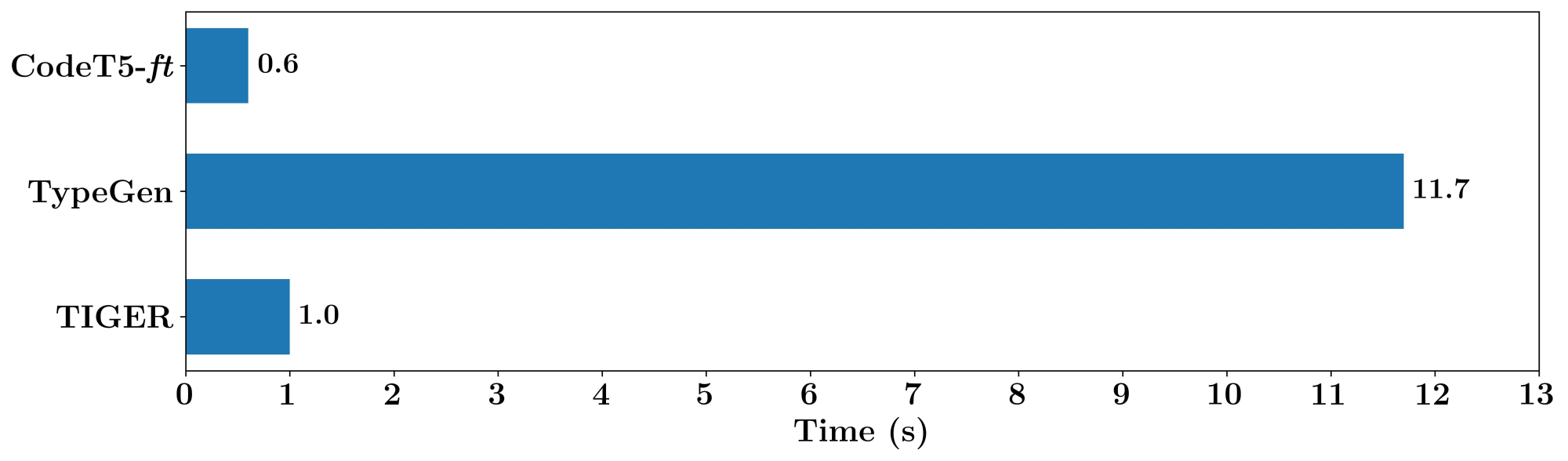}
    \caption{Average inference time of CodeT5-\emph{ft}, TypeGen, and \app.}
    \label{fig:time-plot}
\end{figure}

The average inference time of the three approaches for each instance is depicted in Figure~\ref{fig:time-plot}. Notably, our approach, \app, and CodeT5-\emph{ft} can process one instance within 1 second on average, while TypeGen requires significantly more time for inference (more than 10 seconds). This discrepancy can be attributed to the fact that \app and CodeT5-\emph{ft} employ relatively smaller models compared to TypeGen (220M parameters vs. 175B parameters) and do not rely on complex program analysis such as code slicing. Furthermore, both \app and CodeT5-\emph{ft} have the potential for further optimization in large-scale inference scenarios by leveraging parallel processing on GPUs (\eg increasing the batch size). However, TypeGen faces scalability challenges due to its reliance on OpenAI's online APIs for accurate inference, which significantly impacts the efficiency of TypeGen, given potential issues such as unstable network connections, API request frequency limits, and high financial costs.

\summary{
On average, our approach, \app, processes each instance in approximately 1 second, a performance comparable to CodeT5-\emph{ft} and significantly more efficient than the LLM-based TypeGen. Moreover, \app's efficiency can be further enhanced for large-scale inference scenarios through GPU parallel processing.
}
\subsection{Ablation Study (RQ4)}
We conduct an ablation study to explore the contributions of the two stages of \app.

Table~\ref{tab:two-stage} showcases the inference results of \app and its variants: \textbf{only \emph{generating}} and \textbf{only \emph{ranking}}. When solely utilizing the generating stage, the inference achieves robust accuracy for most type and variable categories, except for (unseen) user-defined types (Column ``Usr (Unseen)''). Conversely, when only employing the ranking stage, the inference demonstrates the highest accuracy for (unseen) user-defined types but performs poorly for the other type and variable categories. By synergizing the strengths of both stages, \app harnesses the creativity of the generation model and the discriminative ability of the similarity model, resulting in consistently practical inference accuracy across all type and variable categories.

\begin{table}
    \caption{Contributions of the two stages. The \first{best} and \second{second-best} results for each category are are highlighted.}\label{tab:two-stage}
    \centering
    \renewcommand{\arraystretch}{1.5}
    \setlength{\tabcolsep}{2pt}
    \begin{threeparttable}
        \begin{tabular}{ccccccccccc}
            \Xhline{2\arrayrulewidth}
            \multirow{2}{*}{Ablation} &  & \multicolumn{9}{c}{\textbf{Top-5 Exact Match (\%)}}                                                                                        \\ \cline{3-5} \cline{7-9} \cline{11-11} 
                                      &  & Ele          & Par          & Usr (Unseen)                  &  & Var           & Arg           & Ret           &  & All           \\ 
            \Xhline{1.5\arrayrulewidth}
            \app                      &  & \first{97.7} & \first{89.4} & \second{94.1} (\second{85.0}) &  & \first{96.2}  & \first{91.6}  & \first{89.1}  &  & \first{94.6}  \\
            \hline
            only \textit{generating}      &  & \first{97.7} & \first{89.4} & \other{80.8} (\other{67.7})   &  & \second{95.3} & \second{82.7} & \second{85.8} &  & \second{91.9} \\
            only \textit{ranking}   &  & 0.7\tnote{*}         & 5.3\tnote{*}         & \first{94.4} (\first{86.5})   &  & 20.0          & 31.8          & 23.4          &  & 21.2          \\ 
            \Xhline{2\arrayrulewidth}
        \end{tabular}
        \begin{tablenotes}
            \item[*] \footnotesize{The non-zero values arise from certain user-defined types sharing names with elementary types.}
        \end{tablenotes}
    \end{threeparttable}
\end{table}



\summary{
    The results of the ablation study confirm the significant contributions of both stages, \ie \emph{generating} and \emph{ranking}, to robust and practical type inference. 
}

\subsection{Case Study}
\begin{figure}
    \centering
    \subfigure[Example 1: Predictions for Missing Type \inlinecode{promise}]{
        \includegraphics[width=1\columnwidth]{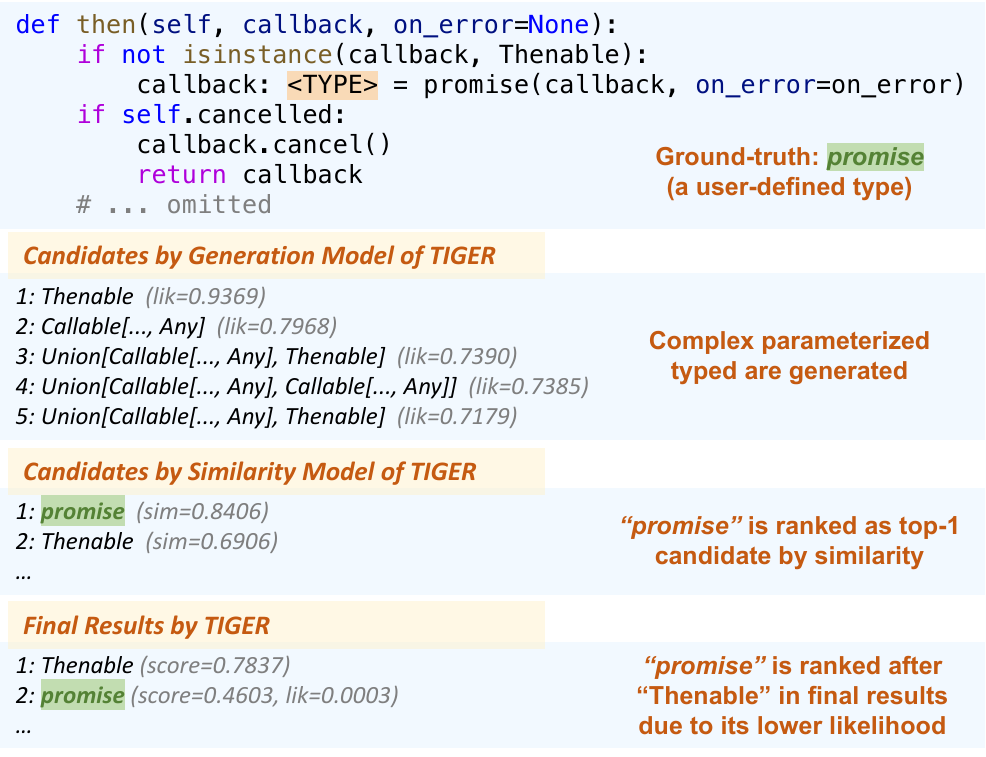}
        \label{fig:case1}
    }
    \subfigure[Example 2: Predictions for Missing Type \inlinecode{Data}]{
        \includegraphics[width=1\columnwidth]{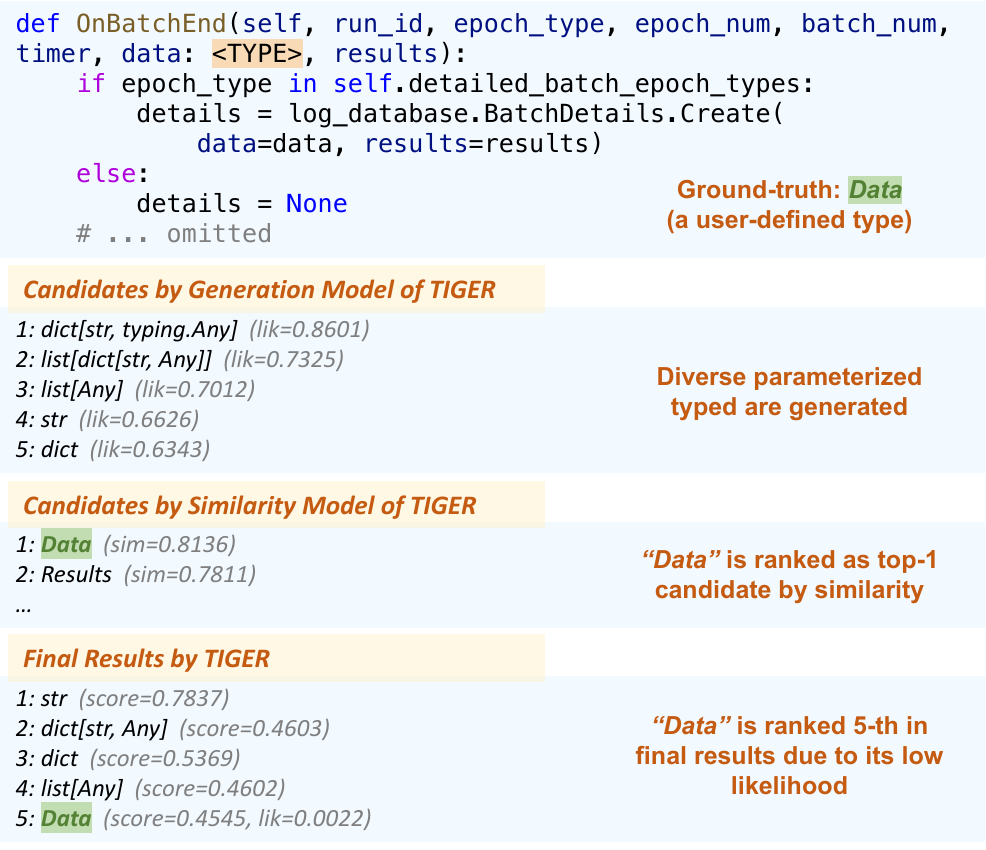}
        \label{fig:case2}
    }
    \caption{Two Suboptimal Cases}
    \label{fig:case-study}
\end{figure}

We have identified certain suboptimal cases, as illustrated in Figure~\ref{fig:case-study}, concerning user-defined types.

In Figure~\ref{fig:case1}, \app's generation model produces complex parameterized types for a type-missed function, with their corresponding generative likelihoods (\ie $lik$). However, the expected type \inlinecode{promise} does not appear in the candidate list. Conversely, \app's similarity model ranks the ground-truth type as the top-1 candidate with a high similarity score (\ie $sim$). When the two models are combined, \app eventually returns \inlinecode{promise} as the 2nd result, just after \inlinecode{Thenable}. Similarly, in Figure~\ref{fig:case2}, the expected \inlinecode{Data} type is absent from the candidates generated by the generation model but is ranked 1st by the similarity model. However, in the final results, it only appears in 5th place.

These suboptimal outcomes can be attributed to the fact that some user-defined types, which are not generated as candidates by the generation model, tend to have lower likelihoods compared to model-generated candidates. As a result, even with high similarities from the similarity model, their final scores, as calculated by Eq.~\ref{eq:final-score}, remain lower. While we have attempted to adjust the weights of the two measures in Eq.~\ref{eq:final-score}, finding an optimal balance has proven challenging. In the future, we plan to explore more effective methods for combining these two measures, such as incorporating likelihood into the contrastive loss.

\section{Threats to Validity}\label{sec:threats}
The primary internal threat to validity arises from potential data errors within the training and test sets. To mitigate this threat, we employ the widely-used ManyTypes4Py dataset for training and evaluating Python type inference approaches, ensuring consistency with common practices in the field. 
Another potential internal threat to validity pertains to the settings of model training hyperparameters. To address this concern, we adhere to standard values for training hyperparameters, including learning rate and epoch numbers. 

Regarding the external threat, although we have consdiered a varity of baselines, the absence of a comparison with DLInfer~\cite{icse/YanFFX23} in our evaluation also poses a external threat to validity. The lack of a publicly available implementation and the complexity of reimplementation make it challenging to include DLInfer for a fair comparison. Instead, we compared with the more recent state-of-the-art approach, TypeGen, which also leverages code slicing similar to DLInfer.

\section{Conclusion and Future Work}\label{sec:conclusion}
This paper introduces \app, a novel Python type inference approach employing a \gtr framework. \app trains a generation model with a generative span masking objective and a similarity model with a contrastive training objective. Both models contribute to the \gtr inference process, generating and ranking candidate types alongside user-defined types based on generative likelihood and contextual similarity. Evaluation on the ManyTypes4Py dataset reveals that \app performs effectively across various type categories, notably demonstrating significant improvements in (unseen) user-defined types. Furthermore, the evaluation results underscore the resilience and efficiency of \app, underscoring the significance of the employed two-stage approach.

In the future, we plan to extend the applicability of \app to other dynamic programming languages such as JavaScript and explore the further integration with static analysis for type validation. Additionally, we aim to adapt \app to address other practical type inference challenges, including those related to partial Python functions.

\section*{Data Availability}
Our replication package is publically available at~\cite{replication}.

\section*{Acknowledgement}
This research / project is supported by the National Research Foundation, Singapore, and the Cyber Security Agency under its National Cybersecurity R\&D Programme (NCRP25-P04-TAICeN). Any opinions, findings and conclusions or recommendations expressed in this material are those of the author(s) and do not reflect the views of National Research Foundation, Singapore and Cyber Security Agency of Singapore.


\bibliographystyle{IEEEtran}
\balance
\bibliography{src/citations.bib}

\end{document}